\def\etal{et~al.}

\newcommand\iso[2]{$^{\rm #1}$#2}

\def\eg{\mbox{e.g.}}
\def\etal{\mbox{\rm et al.}}

\def\rpro{\mbox{$r$-process}}
\def\spro{\mbox{$s$-process}}
\def\ncap{\mbox{$n$-capture}}
\def\teff{\mbox{T$_{\rm eff}$}}
\def\logg{\mbox{log~{\it g}}}
\def\vmicro{\mbox{$\xi_{\rm t}$}}

\documentclass[article]{gwp80}
\usepackage{graphicx}
\usepackage{amssymb}
\tabletypesize{\normalsize}  

\def\plotone#1{\centering \leavevmode
\includegraphics[width=.95\columnwidth]{#1}}

\def\plotone#1{\centering \leavevmode
\includegraphics[width=.95\columnwidth]{#1}}

\shortauthors{Sneden, For, \& Preston}
\shorttitle{RR Lyrae Chemical Compositions}

\begin{document}
\large    
\pagenumbering{arabic}
\setcounter{page}{196}

\title{New Initiatives on RR Lyrae Chemical \\ \\ Compositions}

%
%
\author{{\noindent Christopher Sneden{$^{\rm 1}$},
Bi-Qing For{$^{\rm 1,2}$}, and
George W. Preston{$^{\rm 3}$ }\\
\\
{
\it (1) Department of Astronomy, University of Texas, Austin TX, USA \\
\it (2) ICRAR, University of Western Australia, Crawley WA, Australia \\
\it (3) Observatories of the Carnegie Institution of Washington, 
Pasadena, CA, USA}
}
}

%
%
\email{
(1) chris@verdi.as.utexas.edu 
(2) biqing@astro.as.utexas.edu
(3) gwp@obs.carnegiescience.edu}


\begin{abstract}
The serendipitous discovery by Preston and colleagues of the 
neutron-capture-enhanced RR~Lyrae variable star TY~Gru (a.k.a. 
CS~22881-071 in the ``HK'' survey of very metal-poor halo stars) 
has resulted in a growing set of initiatives on the 
chemical compositions of RR Lyrae stars and their application to 
broader topics in Galactic halo structure. 
Here we summarize the main aspects of our work on TY~Gru,
including a new discussion of our search for possible orbital
motion of this star around a putative unseen companion.
Then we describe a few of the results of a newly-completed intensive 
spectroscopic investigation of 10 additional field RR~Lyr stars.
We finish by outlining current projects that seek to contrast the
atmospheres and chemical compositions of RRc stars with those of
the RRab stars, and that employ a much larger RRab sample in
a chemo-dynamical study of Galactic halo RR~Lyr. 
\end{abstract}

\section{Introduction}

In 2005, the first author of this paper was contentedly continuing 
his investigations into the the contents of neutron-capture elements 
in very metal-poor stars (\eg, Sneden \etal\ 2003; Lawler \etal\ 2004).
At the same time, the second author was beginning graduate study, hoping 
to expand on her initial research into pulsating subdwarf B stars 
(\eg, For \etal\ 2006).
Neither one paid much attention to clear signals that the third author 
was re-kindling his interest in the research that had dominated
his early career (\eg, Preston 1959):  RR~Lyr variable stars.
But a special project a few years ago on the peculiar chemical 
composition of a single, accidentally-rediscovered very metal-poor 
RR~Lyr has expanded into a large-scale general investigation on the 
chemistry of the Galactic RR~Lyr population. 
In this effort, the first two authors have been the happy recipients of 
the data acquisition, the spectrum analyses, the variable star insights, 
and the overall enthusiasm for this work by the third author.

In this contribution we first will review the unusual chemical 
composition of TY~Gru, the star that started us down this research path.  
As part of the TY~Gru discussion we will comment on our continued failure 
to detect orbital motion of TY~Gru around an AGB relic presumed to
be the origin of its abundance anomalies.
Then we will describe the recently-completed detailed investigation
into the atmospheric parameters and abundances of a sample of
10 intensively-observed RRab stars. 
Finally, we will sketch new initiatives that we are pursuing in 
contrasting the spectra of RRc with RRab variables, and in utilizing a 
much larger sample of RRab spectra gathered for the Galactic 
structure studies by Juna Kollmeier and her colleagues.

\begin{figure*}
\centering
\plotone{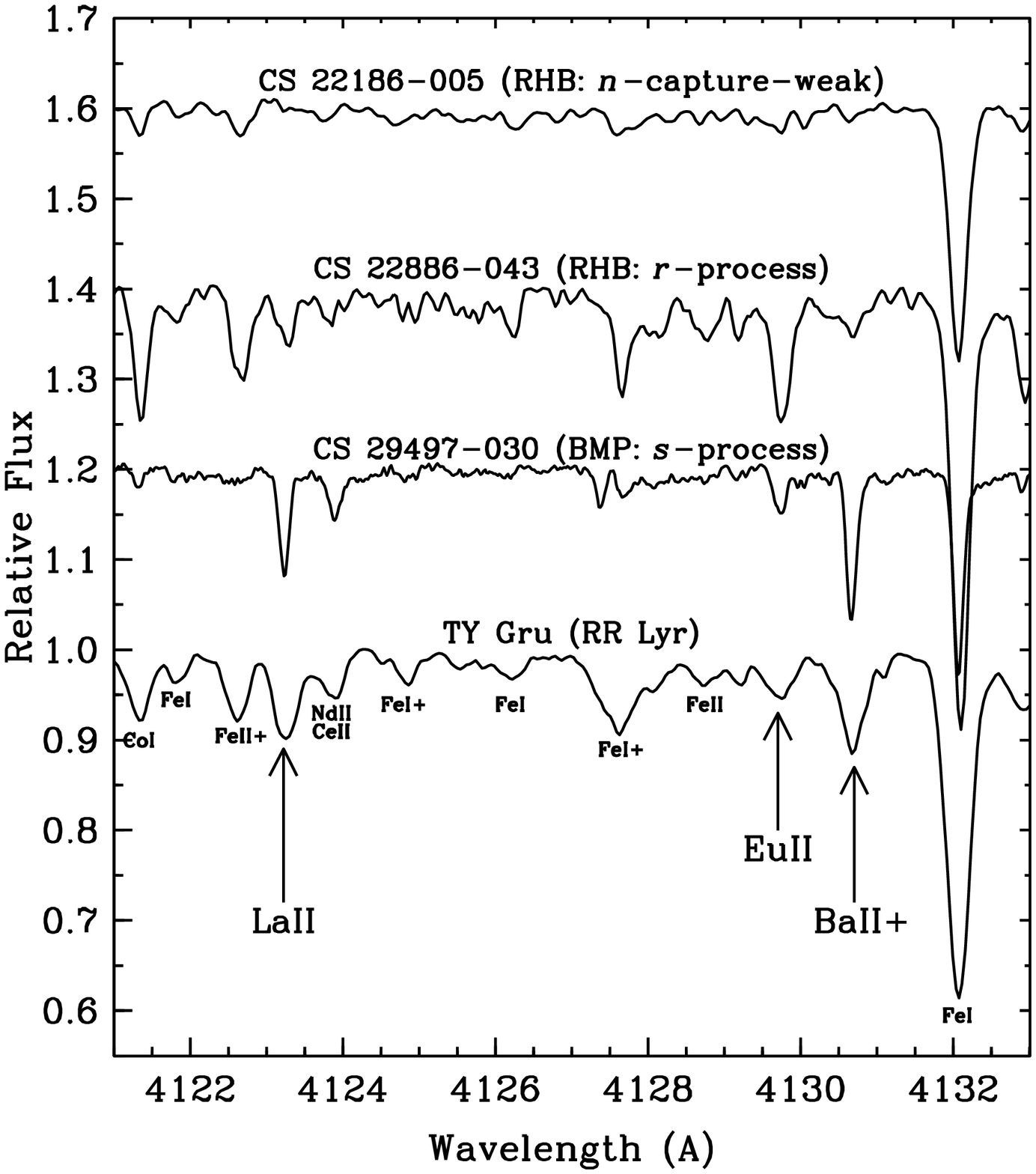}
\vskip0pt
\caption{Figure 1 of Preston \etal\ 2006b:  spectra of four warm 
metal-poor stars in the region of the \ncap\ transitions \ion{La}{2} 
4123~\AA, \ion{Eu}{2} 4129~\AA, and \ion{Ba}{2} 4130~\AA.
The top spectrum is the RHB star CS~22186-005 (Preston \etal\ 2006a), 
which has no enhancements of \ncap\ elements.  
The second spectrum is the RHB star CS~22886-043 (Preston \etal\ 2006a), 
which has an \rpro\ mix of \ncap\ enhancements (\eg, [Eu/Fe]~=~+0.8,
[Eu/Ba]~=~+0.6).  
The third spectrum is the main-sequence blue metal-poor star 
CS~29497-030 (Ivans \etal\ 2005), which has an \spro\  set of 
\ncap\ overabundances (\eg, [Eu/Fe]~=~+2.0, [Eu/Ba]~=~--0.3).  
The bottom spectrum is that of TY~Gru.
\label{tygruf1}}
\end{figure*}

\section{Neutron-Capture Elements in TY~Gruis}

A high-resolution spectroscopic survey (Preston \etal\ 2006a) of 
metal-poor Red Horizontal Branch (RHB) stars discovered in the HK survey 
(Beers, Preston, \& Shectman 1992) included the star CS~22881-071.
Velocity shifts in the two initial observations of this star 
quickly led to recognition that it was the previously identified RRab 
variable star TY~Gru.  
This star was severed from the general RHB survey, and its subsequent 
analysis was published in a separate study (Preston \etal\ 2006b).

We became interested in TY~Gru after noticing that several features
of neutron-capture elements (\ncap, Z~$>$~30) were anomalously
strong compared to RHB stars with similar atmospheric parameters.
In Figure~\ref{tygruf1} we show a small spectral region in the
blue of TY~Gru and some comparison stars.
A practiced eye can see the main spectroscopic results without
any detailed analysis: 
{\bf [1]} all of the \ncap\ features are enhanced compared to those
of an ordinary warm RHB star like CS~22186-005; 
{\bf [2]} the relative strengths of the \ion{Ba}{2} and \ion{La}{2}
lines in TY~Gru are larger than those of \ion{Eu}{2};
{\bf [3]} the TY~Gru \ncap\ line strength pattern is consistent
with that of CS~29497-030, which exhibits products of slow 
\ncap\ synthesis (the \spro) but is incompatible with that
of CS~22886-043, which has a distinct abundance signature of
rapid \ncap\ synthesis (the \rpro).

\begin{figure*}
\centering
\plotone{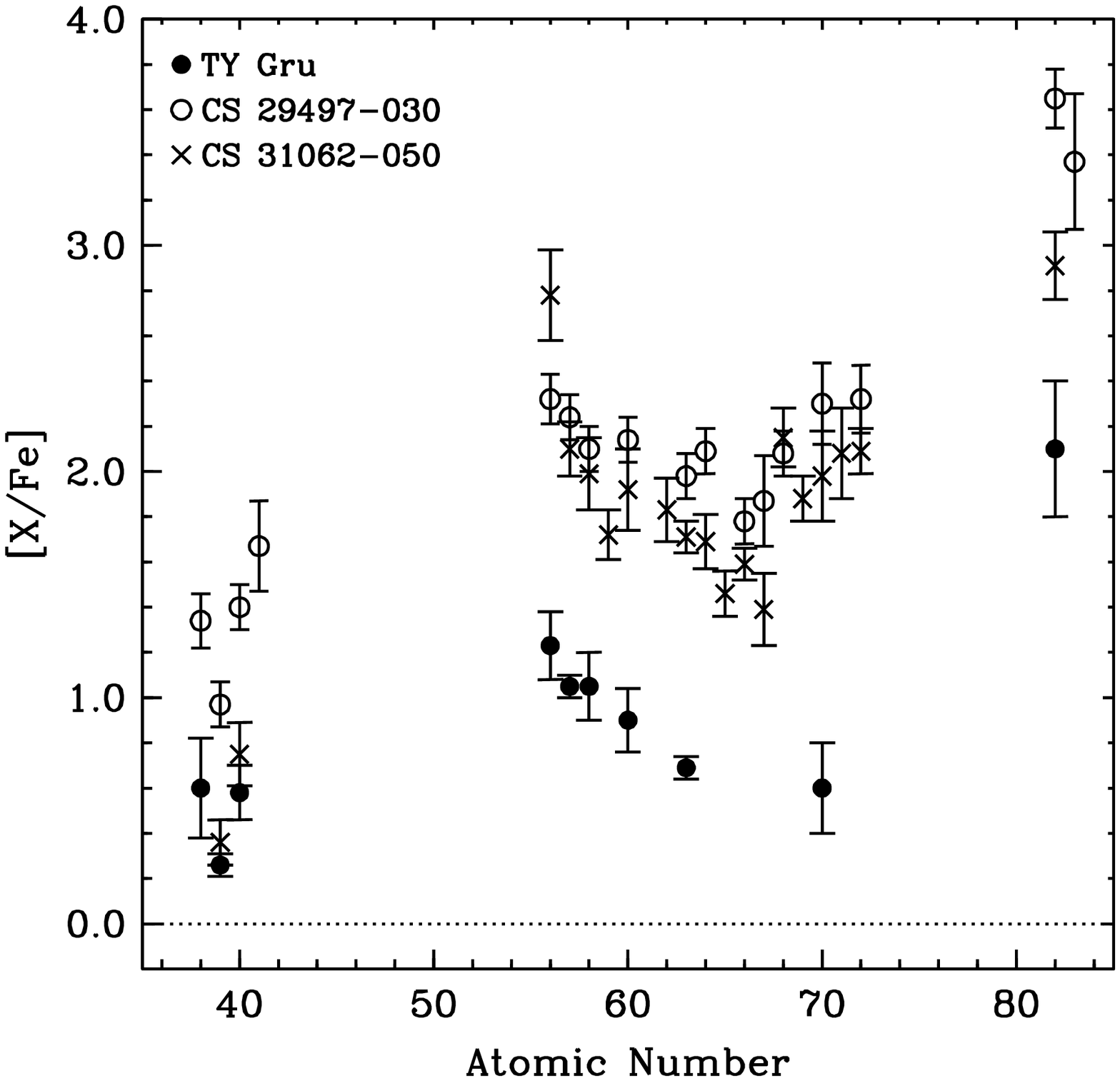}
\vskip0pt
\caption{Figure 9 of Preston \etal\ 2006b:  relative abundance 
ratios of $n$-capture elements in TY~Gru and in Pb-rich metal-poor 
stars CS~31062-050 (Johnson \& Bolte 2004) and CS~29497-030 
(Ivans \etal\ 2005).
The solar abundance ratio [X/Fe]~=~0 is denoted by a dotted line.
\label{tygruf9}}
\end{figure*}

Our detailed abundance analysis, involving 10 elements representing
all observable atomic \ncap\  mass/number domains, demonstrated that
the TY~Gru pattern is a near-perfect match to the abundance patterns of
other very metal-poor, \spro-rich stars.
We illustrate this agreement in Figure~\ref{tygruf9}.  
While the overall amount of \ncap\ abundance excess is less in 
TY~Gru than in CS~29497-030 (Ivans \etal\ 2005) and CS~31062-050 
(Johnson \& Bolte 2004), the agreement in abundance distribution
would be near-perfect if the TY~Gru numbers were to be renormalized
to match, say, the Ba or La abundances of one of the comparison stars.
In this figure, note especially the large overabundances of all 
\ncap\ elements
($+$0.3~$\lesssim$ [element/Fe]~$\lesssim$ $+$2.1); the decline in 
relative overabundances among the rare-earth elements 
as atomic number increases from Z~= 56 to 70; and finally the
enormous overabundance of Pb, the penultimate stable element.
Preston \etal\ (2006b) also derived large carbon overabundances
in TY~Gru.
These are all unmistakable signatures of \spro\ synthesis in 
low-metallicity He-fusion zones of highly evolved (AGB) stars
(\eg, Gallino \etal\ 1998; Goriely \& Mowlavi 2001).

\begin{figure*}
\centering
\plotone{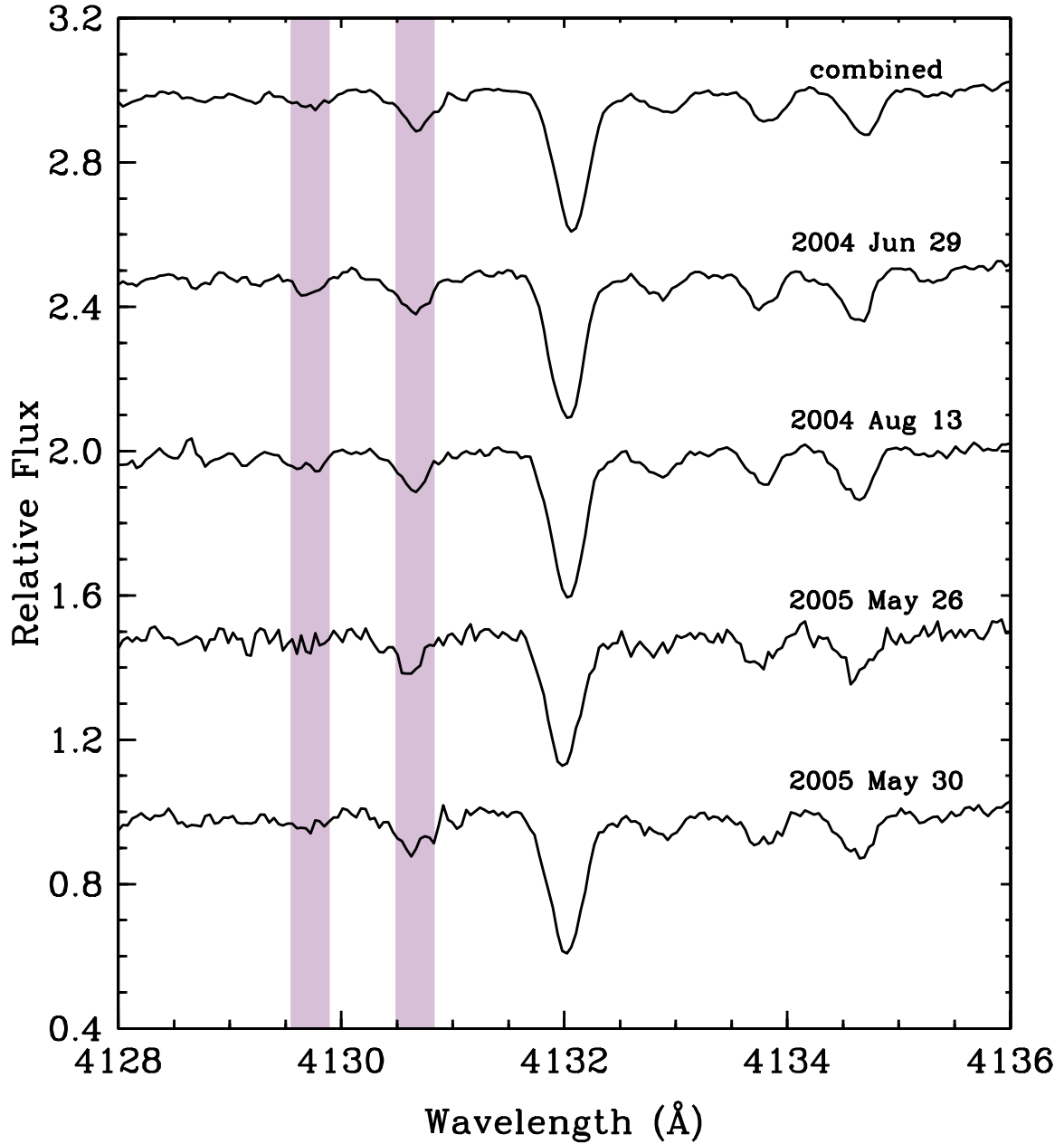}
\vskip0pt
\caption{Adapted from Figure 6 of Preston \etal\ 2006b:  individual 
spectra of TY~Gru taken at phase $<\phi>$~= $+$0.78, and the mean 
``combined'' spectrum (at the top of the figure).
We have added here vertical gray bars to indicate the positions
of the \ion{Eu}{2} transition at 4129.7~\AA\ and the combined
\ion{Ba}{2}$+$\ion{Eu}{2} feature at 4130.6~\AA.
While the individual spectra give little confidence in detection 
of the \ion{Eu}{2} line, it is clearly visible in the combined spectrum.
\label{tygruf6a}}
\end{figure*}

\begin{figure*}
\centering
\plotone{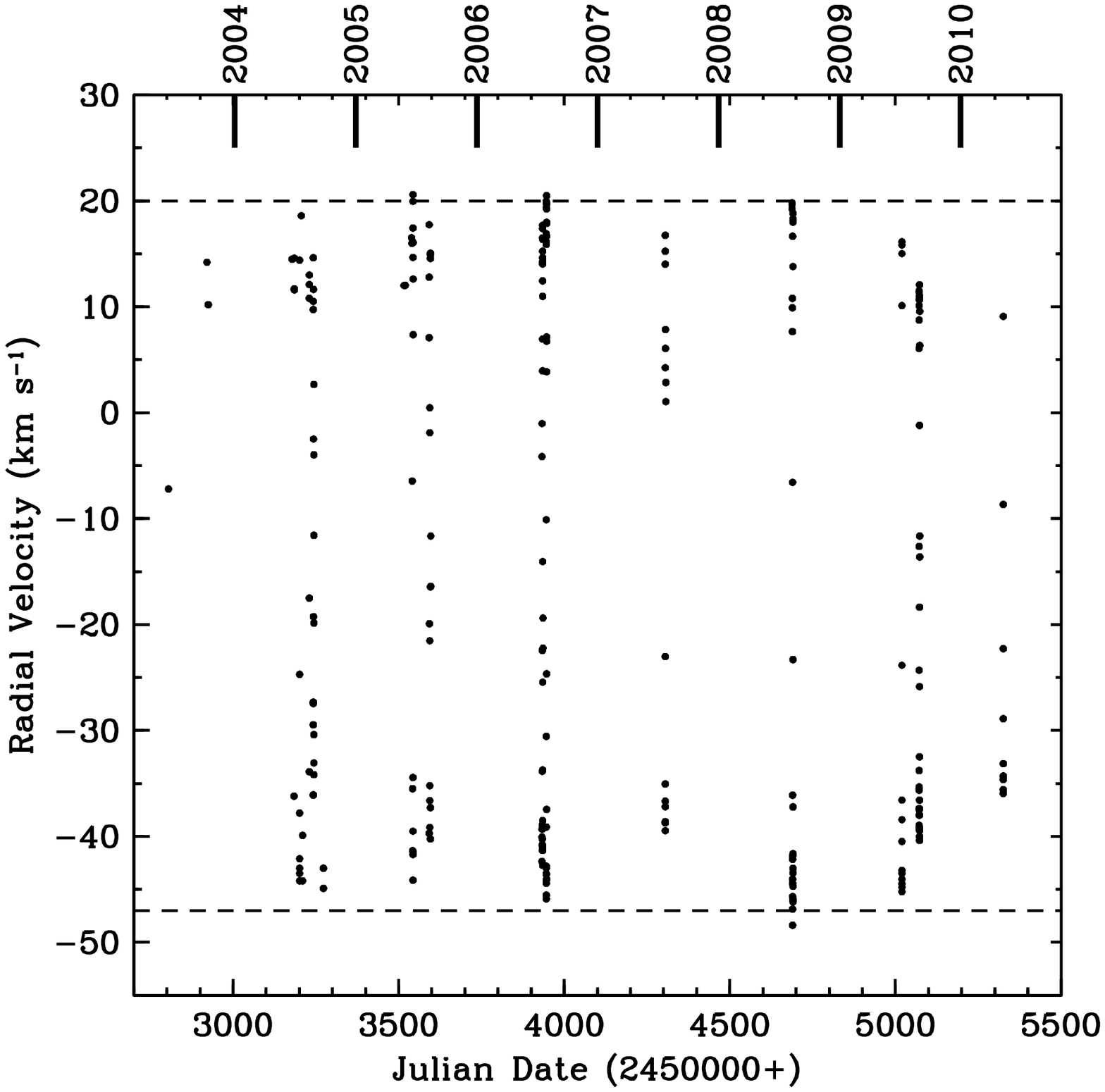}
\vskip0pt
\caption{Radial velocities extracted from individual spectra of TY~Gru 
obtained from 2003 through 2010.
The data are shown as a function of Julian date, but to assist the
reader we have drawn vertical lines at the top of the figure to
denote January~1 of the year that is written at the top.
Horizontal lines have been drawn to mark the radial velocity
envelope limits for TY~Gru.
\label{tygruvel}}
\end{figure*}

The TY~Gru investigation featured an observational/analytical advance in
RR~Lyr spectroscopy which has proven to be important in our recent work.
Much of the previous RR~Lyr abundance work (\eg, Clementini \etal\ 1995; 
Lambert \etal\ 1996) was relatively photon-starved and consequently
employed few spectra.
These data were generally obtained at one mean phase position 
with substantial phase smearing.
Thanks to the availability of substantial amounts of time on 
large telescopes equipped with efficient echelle spectrographs, 
Preston \etal\ (2006b) were able to obtain 82 high-resolution spectra 
in maximum integration times of about 15 minutes, which correspond to a 
pulsational phase interval of $\Delta\phi$~= 0.02. 
This procedure produced individual spectra of relatively low 
signal-to-noise (S/N), but the repeatability of RR~Lyr 
spectroscopic variations allowed us to improve S/N by addition 
several spectra obtained at similar phases, as we illustrate in
Figure~\ref{tygruf6a}.
The phase for the TY~Gru abundance analysis, $\phi$~$\sim$ 0.8,
was chosen to be near light minimum in the belief that this would be
the optimal phase (most stable photospheric conditions).
Although this had been a popular choice in past RR~Lyr studies, our
subsequent work has forced a re-evaluation; see \S\ref{biqingwork}.

TY~Gru is the first RR~Lyr star to be found with large carbon 
and \spro\ abundance enhancements.
Nearly all metal-poor stars with these abundance characteristics
are proven or suspected members of binary star systems with unseen 
secondary companions (\eg, McClure 1997; Preston \& Sneden 2001; 
Lucatello \etal\ 2005).
This is the simplest explanation for the TY~Gru abundances, but
observational proof of the idea has been difficult to obtain.

Preston \etal\ (2006b) were unable to detect any secular drift in
the the TY~Gru systemic radial velocities that might be indicative
of wide-binary orbital motion.
At that time however only two years of spectra contributed to the
radial velocity information.
In subsequent years more spectra have been gathered, and Preston's (2011)
Figure~1 displays the expanded velocity set. 
Even more recent TY~Gru spectra leads to one more update which
we show in Figure~\ref{tygruvel}.
TY Gru is a Blazhko star which exhibits a modest variation of 
radial velocity amplitude in a period of about 68 days as 
illustrated in Figures~10 and 11 of Preston \etal\ (2006b).  
The Blazhko phenomenon produces a change in velocity amplitude 
but not in the systemic velocity. 
In Figure~\ref{tygruvel} we see such amplitude variations, but the 
minimum and maximum radial velocities in each observing season 
lie within the dashed horizontal lines in the figure; they do not 
appear to have shifted up or down during the past seven years.

This lack of secular velocity drift for TY~Gru can be used to
place a constraint on the period of orbital motion.
We adopt a mass for TY~Gru of M~=~0.68M$_{\odot}$, a typical value
for RR~Lyr stars (Castellani, Castellani, \& Cassisi 2005),
and assume a mass for a putative white-dwarf companion of
M~$\approx$~0.6M$_{\odot}$.
From the observed dispersions in upper and lower bounds of the radial
velocities shown in Figure~\ref{tygruvel}  we believe that we could detect a 
change in systemic velocity of 5~km~s$^{-1}$ during the past 
2000 days (half orbital period) had it occurred. 
The circular velocities for our adopted masses and orbital periods of 
4000 and 8000 days are 6.8~km~s$^{-1}$ and 5.4~km~s$^{-1}$, respectively.  
These periods bracket the longest orbital periods (5324~d and 6489~d) 
that have been found for marginally detectable Ba/CH giants 
(Boffin \& Za\u{c}s 1994).  
From the above considerations we argue that the orbital period of 
the putative TY Gru binary must be substantially greater than 4000 d.  
There must be a longest period (largest binary separation) for which 
wind accretion from an AGB companion will fail to produce the 
observable pollution of the TY Gru envelope.   
Thus, the search for orbital motion of TY Gru has interesting 
astrophysical implications.  
We note, finally, that Preston \& Sneden (2001), and Preston (2009b) 
have reported similar difficulty in the 
detection of orbital motion in several main sequence carbon stars.

\section{An Intensive Spectroscopic Investigation of Field RRab 
Variables\label{biqingwork}}

A larger-scale investigation of RR~Lyr was conceived in part to
understand if possible binary mass transfer from a retired 
compact-star companion had created a unique evolutionary HB status
for TY~Gru (Preston 2011).
A sample of 10 RRab stars with photometric properties similar to TY~Gru
were intensively observed with the Las Campanas du~Pont telescope and
its echelle spectrograph; extant spectra of TY~Gru were added in.
The resulting data set consists of over 2300 short-integration spectra, 
about 200 spectra/star, covering the spectral range 
3500~\AA~$<$ $\lambda$~$<$ 9000~\AA, with resolving power 
$R$~$\sim$ 27,000, and typical S/N of 20 in the blue.
All phase intervals of all program stars have been sampled with multiple
spectra.

This rich database can be mined to address a variety of questions
about RR~Lyr atmospheric physics, and already Preston (2009a) has
discovered unexpected helium emission lines at some phases in
all of the stars.
We also have used these spectra to refine their velocity information 
and pulsational ephemerides (For, Preston, \& Sneden 2011a).

Here we outline just a few aspects of our derivation of $photospheric$ 
parameters \teff, \logg, and \vmicro, overall [Fe/H] metallicities, and 
relative abundance ratios [X/Fe] for our RRab sample.
Detailed results are described in For, Sneden, \& Preston (2011b).

\begin{figure*}
\centering
\plotone{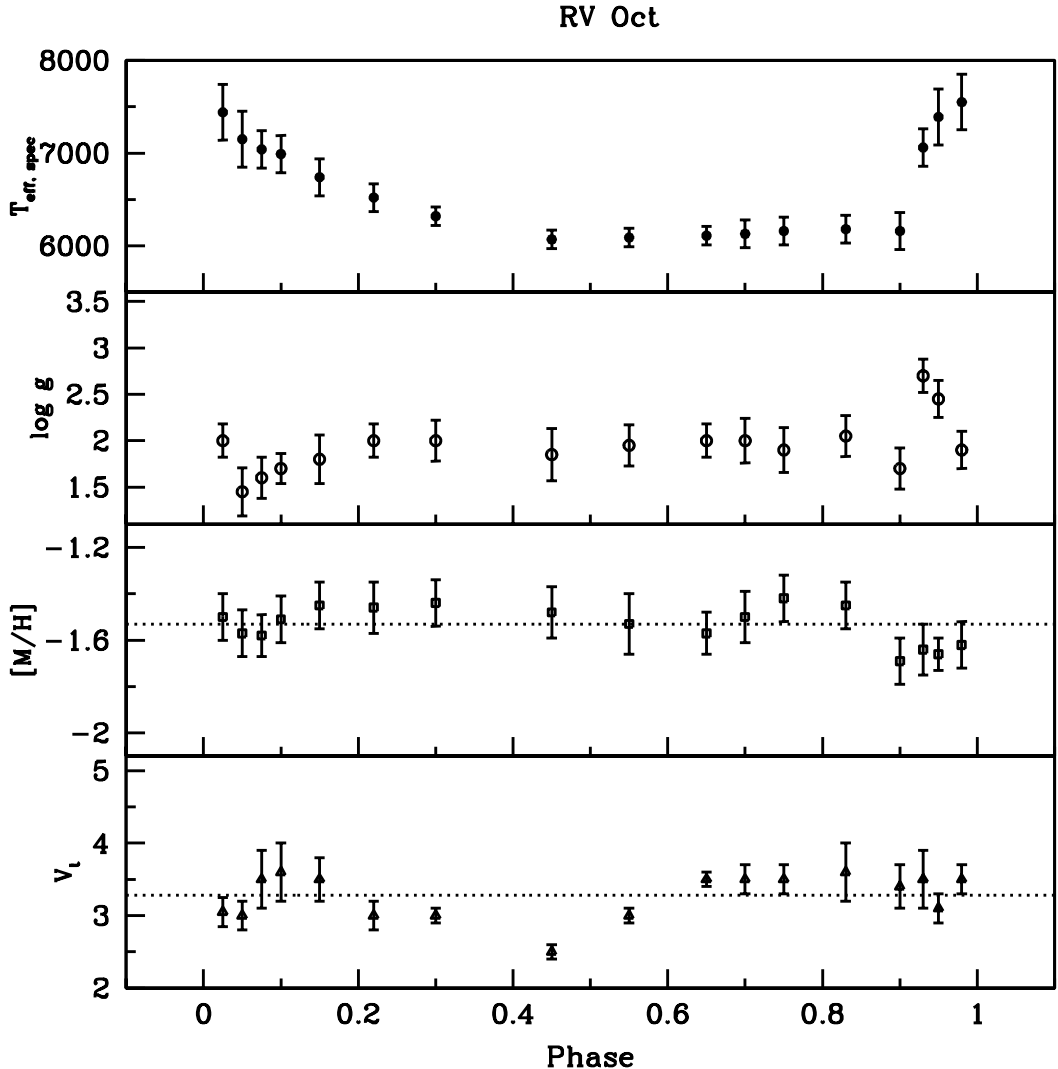}
\vskip0pt
\caption{Variations in atmospheric parameters with pulsational phase
for RV~Oct.
The $>$200 individual spectra for this star have been binned into
17 phase intervals to increase S/N.
\label{rvoct1}}
\end{figure*}

\begin{figure*}
\centering
\includegraphics[width=.65\columnwidth]{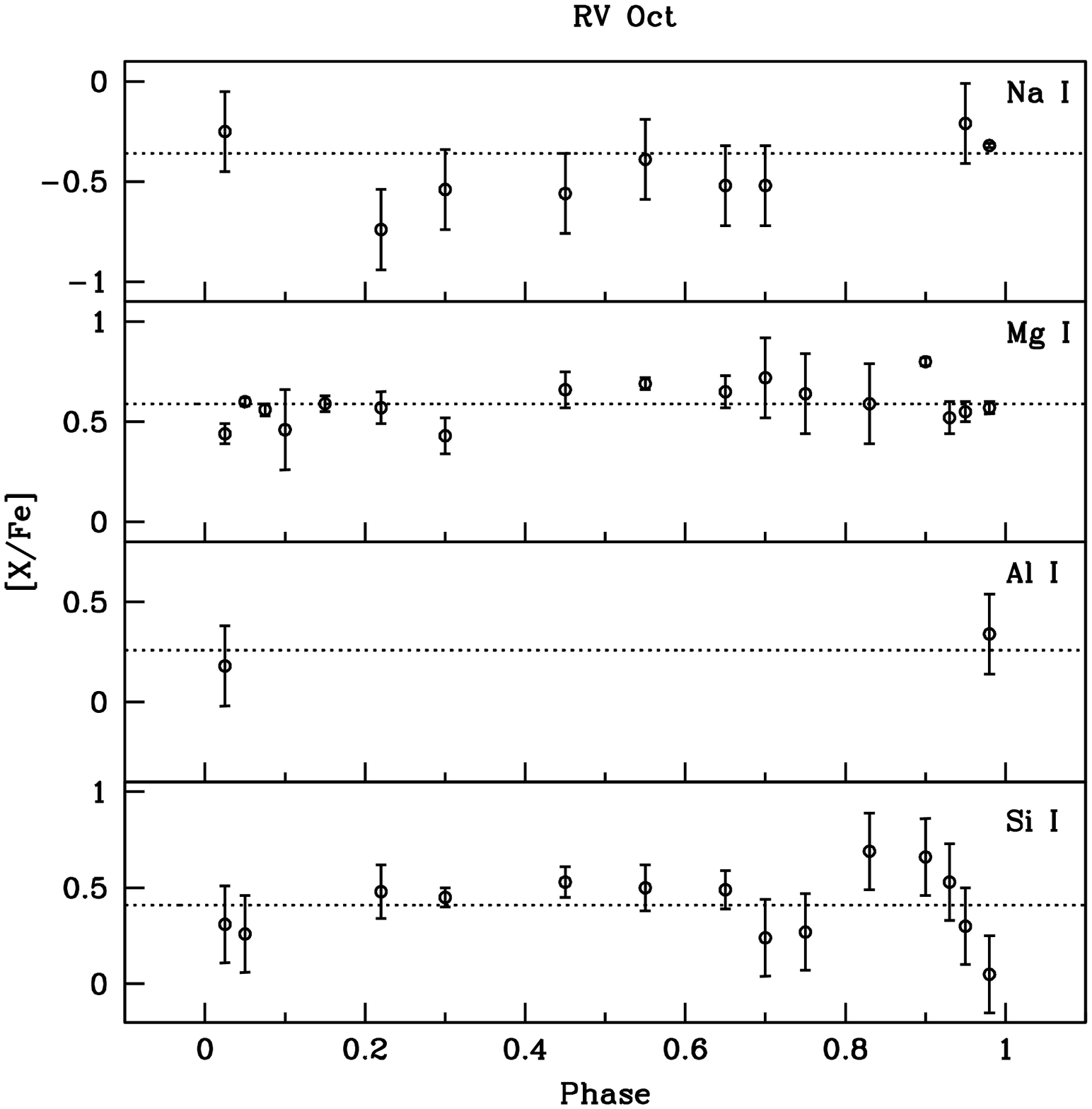} \vfil
\includegraphics[width=.65\columnwidth]{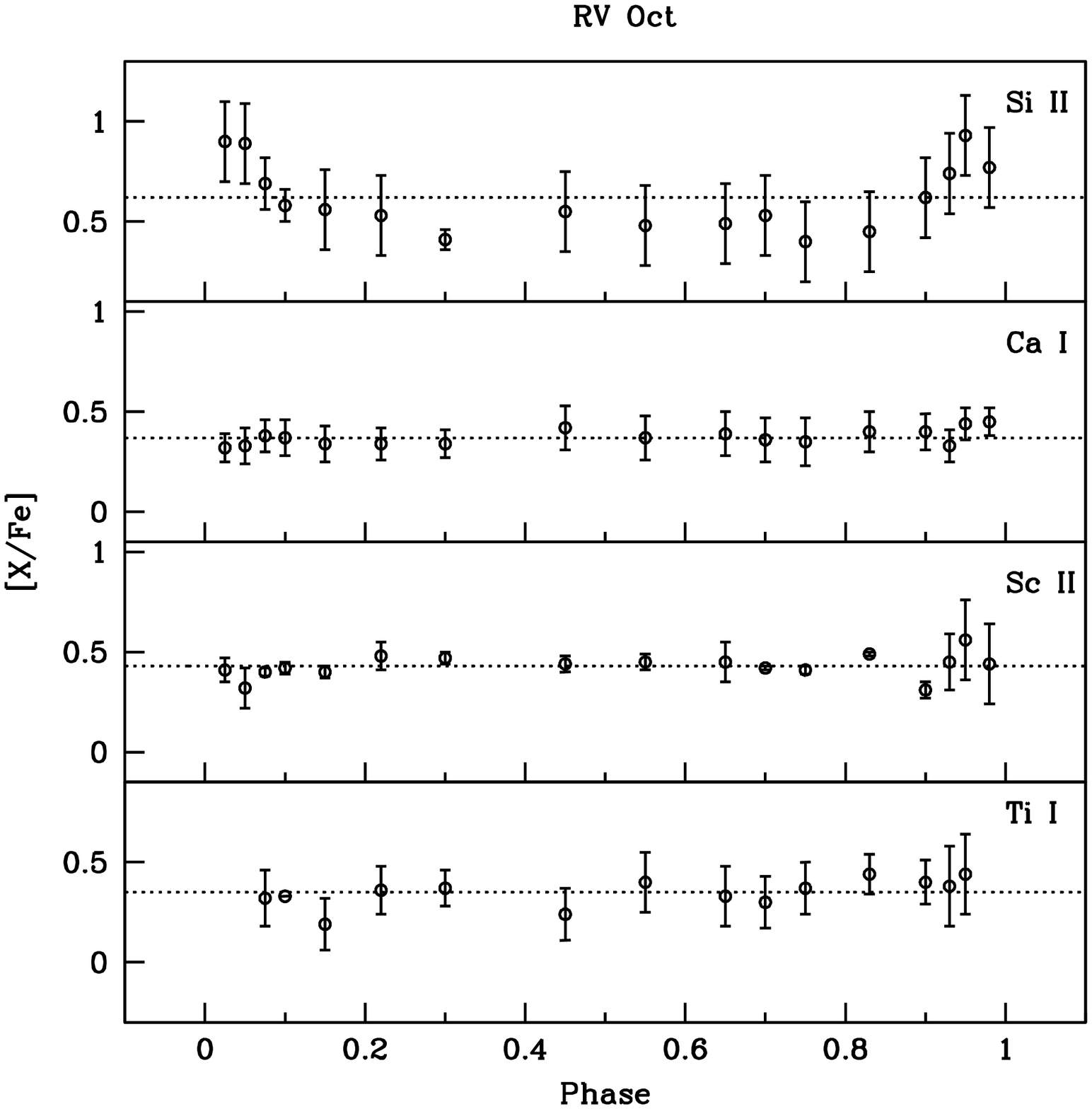}
\vskip0pt
\caption{Variations of the relative abundances of some light elements 
in RV~Oct.
\label{rvoct2}}
\end{figure*}

\begin{itemize}

\item Temperature, gravity, and microturbulence vary in similar 
and predictable ways throughout the pulsational cycles 
in all of our RRab sample.
As one example, in Figure~\ref{rvoct1} we show the variations in 
these quantities in program star RV~Oct.
The phase-dependent atmospheric parameters are for a particular
RRab variable, but pulsational phases changes are qualitatively
always the same in our program  stars (and most likely in all
``normal'' RRab stars as well).

\item The optimal phase for abundance studies, defined as that
phase at which the metal absorption lines are sharpest and most symmetric,
occurs at $\phi$~$\sim$ 0.35, not at the mimimum light phase of
$\phi$~$\sim$ 0.75.
Future RR~Lyr chemical composition studies should concentrate
on $\phi$~$\sim$ 0.35 if possible.

\item In spite of \teff\ changes of more than 1000~K and \logg\
changes of more than 1~dex during each pulsational cycle,
the [Fe/H] metallicities displayed in Figure~\ref{rvoct2} appear
to be essentially invariant to within the uncertainties of an individual
abundance determination, over the entire RV~Oct pulsational cycle.
This statement can be applied to our whole sample of RRab stars,
suggesting that reliable metallicities can be determined at
any RR~Lyr phase, as long as good estimates of \teff, \logg, and
\vmicro\ can also be extracted from the spectra.

\item Just as importantly, nearly all derived relative abundance ratios
are insensitive to pulsational phase, and are mostly in accord
with expectations based on abundance ratios for other halo-star
evolutionary groups.
In Figure~\ref{rvoct2} we show variations of eight light-element abundance
ratios, and one can easily divide the confidence levels of claimed
phase invariance into excellent (\ion{Ca}{1}, \ion{Sc}{2}, \ion{Ti}{1}), 
probable (\ion{Na}{1}, \ion{Mg}{1}), uneasy (\ion{Si}{1}, \ion{Si}{2})
and unproven (\ion{Al}{1}).

\item The elements Si and Ca are official members of the light
$\alpha$-element group, whose major isotopes are multiples
of helium nuclei.
Element Ti has a courtesy membership in the $\alpha$ group, because
while its major isotope \iso{48}{Ti} is not an aggregate of $\alpha$
particles, its abundance in most metal-poor stars mimics those of the 
true $\alpha$ elements.
We find the abundance ratios [Mg, Si, Ca, or Ti/Fe]~$\sim$ +0.5, in
qualitative agreement with the elevated $\alpha$ abundances seen 
in other types of metal-poor stars (\eg, Cayrel \etal\ 2004).  

\item Si abundances based on \ion{Si}{1} lines in the blue spectral
region are known to be temperature-sensitive 
(\eg, Preston \etal\ 2006a, Sneden \& Lawler 2008).
As one can see in Figure~\ref{rvoct2}, the addition of \ion{Si}{2} lines
in RV~Oct does little to alter the situation.
Happily however, on average both species support the notion of
Si overabundance in this and in the rest of our RRab sample.

\end{itemize}

These comments only touch on the some of the For \etal\ (2011b)
results.
Please see that paper for extended comments on the above points,
and for discussion of the relative abundances of heavier elements, 
the variation of microturbulent velocity with pulsational phase, 
the relationship between \vmicro\ and line full-width-half-maxima, 
and a new method to estimate RRab effective temperatures without
extended analyses.

\section{Some Current Projects}

\begin{figure*}
\centering
\plotone{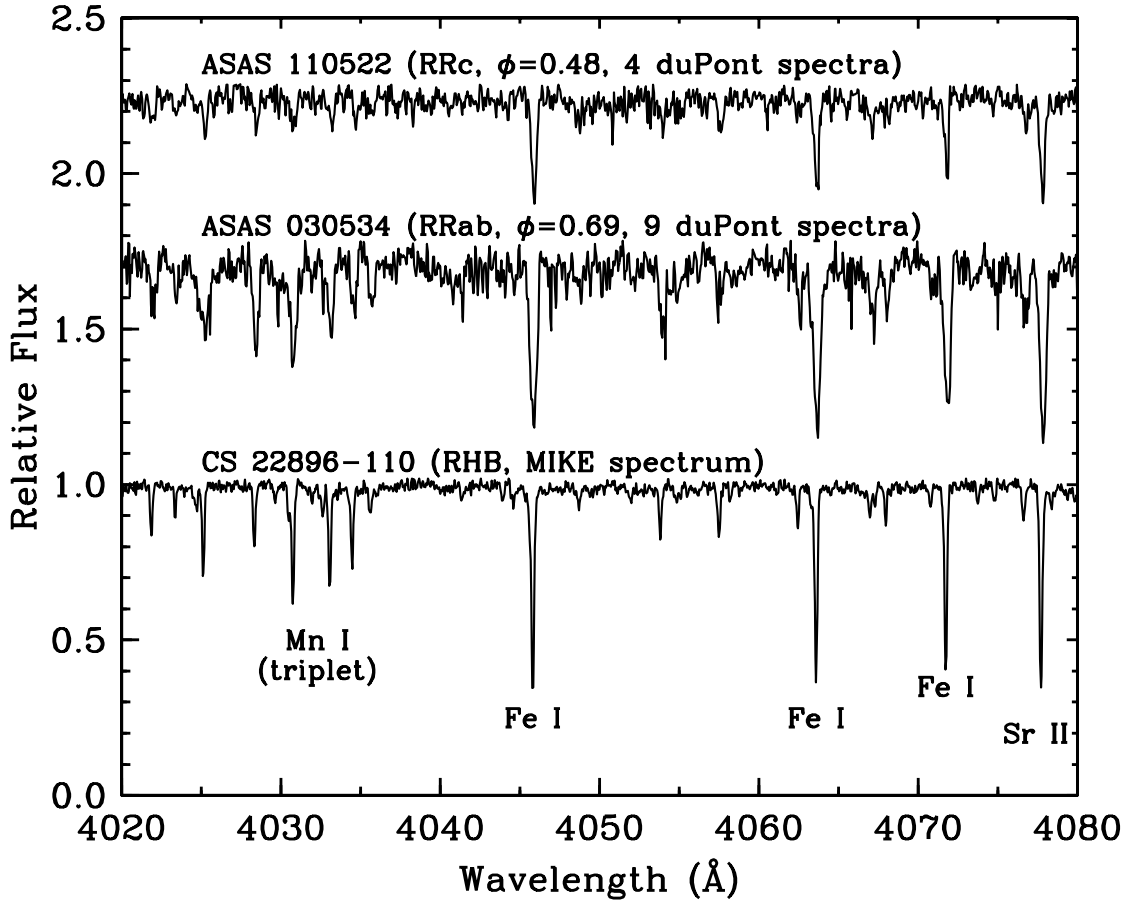}
\vskip0pt
\caption{Example spectra of an RRc variable (top), an RRab (middle)
and a non-variable red horizontal-branch star (bottom).
The single RHB spectrum was obtained with a larger telescope (Magellan)
at high spectral resolution ($R$~$\sim$ 45,000) and with a long
exposure time, while the spectra of the RR~Lyr stars were gathered
with the smaller du Pont telescope, lower resolving power 
($R$~$\sim$ 27,000), and very short exposure times.
The number of RR~Lyr individual spectra that were co-added, and the
mean phases of the spectra, are indicated in the figure legend.
\label{rrcspec2}}
\end{figure*}

Most previous spectroscopic studies of horizontal-branch variables have 
focused on the RRab stars, the majority component of the RR~Lyr population. 
Much less attention has been paid to the RRc stars.  
The RRc stars present some observational challenges. 
They are relatively rare (hence fainter on average), and they are hotter 
so their absorption lines are weaker at constant abundance.

We have begun a program to derive atmospheric parameters and 
elemental abundance ratios in a sample of field RRc variables.
We employ the same du Pont echelle configuration that was used to
obtain the RRab spectra described above. 
In spite of the faintness of the targets we have obtained satisfactory 
data by co-adding individual spectra taken at similar phases. 
In Figure~\ref{rrcspec2} we present a small spectral region in one of 
the RRc target stars.
Obviously the spectra of this star and a comparison RRab shown 
in this figure are not as good as that of an RHB star studied by 
Preston \etal\ (2006a); the smaller telescope and shorter 
spectroscopic exposure times result in lower S/N. 
However, one can clearly see all the major absorption features in 
the RRc spectra that are necessary to determine both overall metallicity 
and enough abundance ratios to be able to describe the chemical history 
of the RRc stars.
Note also from comparison of the top and middle spectra in 
Figure~\ref{rrcspec2} that the FWHM (macroturbulence) of 
the RRc absorption lines is lower than that of the RRab star.

\begin{figure*}
\centering
\plotone{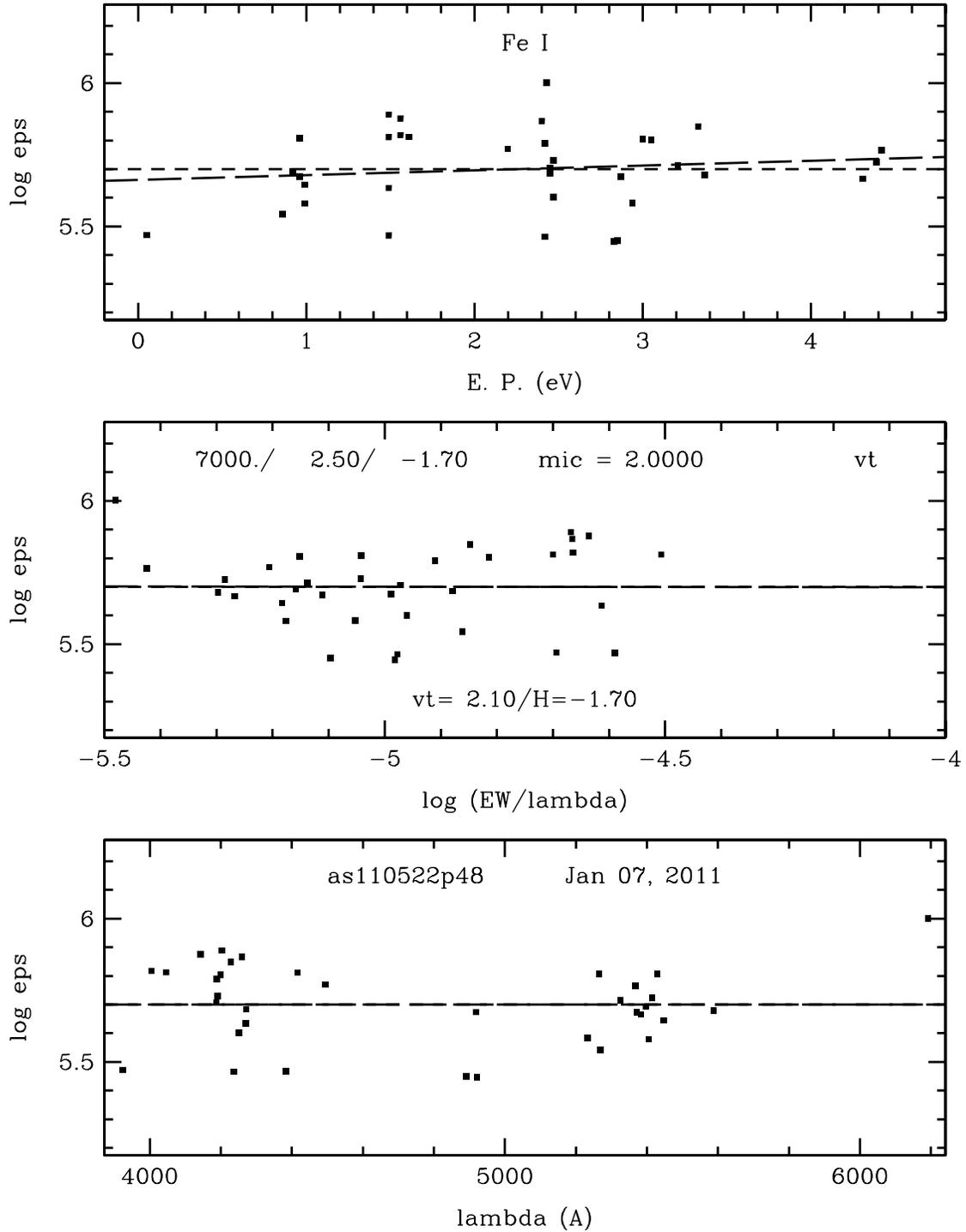}
\vskip0pt
\caption{Abundances of individual \ion{Fe}{1} lines in the RRc
star AS110522 at phase $\phi$~$\approx$ 0.48.
The top panel has the run of these abundances with excitation potential,
the middle panel with the log of the reduced width,
and the bottom panel with wavelength.
\label{rrc}}
\end{figure*}

Analysis of the RRc spectra are underway.
We have obtained preliminary estimates of the atmospheric parameters of one
of the stars. 
To accomplish this, we began with the same Fe input line list that was
used by For \etal\ (2011b), then measured equivalent widths, and performed
a standard atmospheric analysis.  
In Figure~\ref{rrc} we show the abundances of individual \ion{Fe}{1} lines
as functions of excitation energy, line strength, and wavelength.
Not shown in this figure are the abundances from \ion{Fe}{2} lines.
The derived model satisfied the usual spectroscopic criteria of no
substantial trends of \ion{Fe}{1} abundances with excitation potential,
equivalent width, and wavelength, along with equality of the mean
\ion{Fe}{1} and \ion{Fe}{2} abundances.
The temperatures and gravities are consistent with general expectations
for RRc variables.
The derived metallicity, 
[Fe/H]~$\equiv$ log~$\epsilon_{star}$~$-$ log~$\epsilon_{Sun}$~=
5.7 $-$ 7.5 = $-$1.8, clearly indicates membership of this star 
in the Galactic halo population.
But most intriguingly, we derive \vmicro~$\approx$ 2.1~km~s$^{-1}$,
much smaller than typical values of RRab stars (3$-$4~km~s$^{-1}$).
This points to absence of shock-wave disturbances in RRc atmospheres
at the microscopic level.

We also are participating in Juna Kollmeier's statistical parallax
program.
A major spectroscopic survey has been undertaken by Kollmeier and
her colleagues to derive radial velocities and abundances for more than
1000 RR~Lyr stars.
The principal goal of this project is to calibrate the fundamental 
distance scale using the RR~Lyr as standard candles.  
The spectra being gathered for this project have relatively low S/N, 
of necessity.  
The spectra are more than adequate produce good radial velocities.  
Abundance determinations pose more of a challenge. 
Our part of the overall effort will be to derive metallicities and
a limited set of abundance ratios.
We are finding that that reliable metallicities can be obtained 
with spectra of very weak signal by using the ``multiplex'' 
advantage of the many Fe-group element absorption features that are 
present in RR~Lyr stars.
We are employing techniques that are similar to those pioneered
by Carney \etal\ (1987).
We are initially concentrating on the 5000$-$5400~\AA\ spectral region
(as did Carney \etal) which contains the \ion{Mg}{1} b~lines and other
strong spectral features.
For detailed abundance ratios other spectral domains will need
to be surveyed (\eg, the 4500~\AA\ region which contains \ion{Ba}{2}
4554~\AA).
Observations for this program are underway.

\section{Acknowledgments}

We are grateful for support from the U.S. National Science Foundation
over many years, currently with grant AST~0908978.

\end{document}